\documentclass[aps,twocolumn,prl,superscriptaddress,showpacs,showkeys,floatfix]{revtex4-1}

\usepackage{amsbsy} 
\usepackage{graphicx}
\usepackage{amsmath}
\usepackage{amssymb}
\input{epsf}

\begin{document}
\title{Exotic States in the Dynamical Casimir Effect}

\author{Giuliano Benenti}
\affiliation{CNISM and Center for Nonlinear and Complex Systems,
Universit\`a degli Studi dell'Insubria, via Valleggio 11, 22100 Como, Italy}
\affiliation{Istituto Nazionale di Fisica Nucleare, Sezione di Milano,
via Celoria 16, 20133 Milano, Italy}
\author{Stefano Siccardi}
\affiliation{Department of Information Technologies, University of Milan,
via Bramante 65, 26013 Crema, Italy}
\author{Giuliano Strini}
\affiliation{Department of Physics, University of Milan,
via Celoria 16, 20133 Milano, Italy}

\begin{abstract}
We consider the interaction of a qubit with a single mode of
the quantized electromagnetic field and show that, in the 
ultrastrong coupling regime and when the qubit-field interaction
is switched on abruptly, the dynamical Casimir effect leads
to the generation of a variety of 
exotic states of the field, which cannot
be simply described as squeezed states.
Such effect also appears when initially both the qubit and 
the field are in their ground state. 
The non-classicality of the obtained exotic 
states is characterized by means of a parameter based on
the volume of the negative part of the Wigner function. 
A transition to non-classical states is observed by changing
either the interaction strength or the interaction time. 
The observed phenomena appear as a general feature of 
nonadiabatic quantum gates, so that the dynamical Casimir effect
can be the origin of a fundamental upper limit to the 
maximum speed of quantum computation and communication protocols.
\end{abstract}


\maketitle

In 1948 Casimir \cite{casimir} predicted that two 
uncharged conducting parallel plates, 
experience an attractive force. This phenomenon, known
as (static) Casimir effect, is explained in terms of quantum mechanical
vacuum fluctuations of the electromagnetic field. Indeed the two
plates impose boundary conditions to the field such that  
the density of electromagnetic modes between the plates
depends on their distance $d$. The plates affect
the \emph{virtual photons} which constitute the field, 
thus generating a net 
attractive
force proportional to $d^{-4}$. 
The static Casimir effect, 
experimentally demonstrated for the first time by Sparnay in 1958 
\cite{sparnay}, plays a very important role both in 
fundamental physics investigations and in understanding 
the basic limits of nanomechanical technologies, often
with surprising results 
\cite{capasso,lambrecht,lamoreaux}. A similar evolution can be foreseen
for the dynamical Casimir effect (DCE) \cite{moore,dodonov},
for quantum computation and communication protocols.

The DCE concerns the generation of \emph{real photons} from
the vacuum due to time-dependent boundary conditions, for 
instance when the distance $d(t)$ between the 
two plates changes in time. The DCE is 
closely related \cite{noriRMP} to other 
quantum vacuum amplification mechanisms such as the 
Unruh effect \cite{unruh} and the Hawking radiation \cite{hawking}.
The DCE has been recently demonstrated 
experimentally in superconducting circuits \cite{norinature,lahteenmaki},
in the framework of circuit quantum electrodynamics
(circuit-QED) \cite{blais,wallraff}.

In the paradigmatic model of a qubit-oscillator system,
within the rotating wave approximation (RWA) the ground state is the product of 
the qubit's ground state and the oscillator's vacuum state. 
Starting from an initial state with both the qubit and the 
oscillator in their ground states, within the RWA 
the state remains unchanged even when the interaction is turned on.
On the other hand, the inclusion of the interaction terms
beyond RWA leads to squeezed or cat states
containing virtual photons \cite{bastard,noriPRA,grifoni},
which can be released as real photons under abrupt switching of the 
matter-field coupling \cite{bastard,sorba}.
The RWA is a good approximations for a two-level atom (qubit) 
in a resonant cavity (oscillator), where the ratio between the 
frequency $\Omega$ of the Rabi oscillations between the two 
relevant states of the atom and the cavity frequency $\omega$ 
is typically $10^{-6}$ \cite{haroche}. On the other hand, terms
beyond the RWA cannot be neglected in circuit-QED experiments, 
where one can enter the so-called
\emph{ultrastrong coupling regime}, in which the two 
frequencies $\Omega$ and $\omega$ become comparable
\cite{bourassa,gross,mooij}. Such regime is of great interest for
quantum computation, as high-speed operations are 
needed to perform a large number of quantum gates within the 
decoherence time scale, as requested to operate fault tolerantly
\cite{qcbook,nielsen}.   

In this paper we study the dynamics of a qubit-oscillator
system, in the \emph{nonadiabatic regime} in which the coupling 
is switched on and off suddenly. We find that, by varying the
interaction strength and the interaction time, a great variety
of exotic states of the field are generated, both with and
without measurement of the final state of the qubit,
even when in the initial state both the qubit and the oscillator 
are in their ground state.
The non-classicality of such states is characterized by means of 
a negativity parameter based on the volume of the
negative part of the Wigner function \cite{zyczkowski,benedict}. 
We show that such parameter indicates a transition to 
non-classical states by varying either the interaction
strength or the interaction time.

We stress the important differences between the problem 
considered here and the standard QED:
(i) we consider a single mode rather than an infinite number of modes;
(ii) the quantization volume (i.e., the volume of the cavity)
is fixed and the limit of infinite volume is not taken at the end
of the computations;
(iii) while in standard QED the interaction is usually 
switched on adiabatically to avoid transient phenomena, here 
we switch on and off the interaction abruptly and focus
exactly on transient phenomena. 
Moreover, differently from cavity QED, where the quantization volume
is usually
larger than $\lambda^3$, with $\lambda$ wave length of the single
mode, in circuit QED by means of microstrips one 
can achieve quantization volumes much smaller than $\lambda^3$. 
As a consequence, the ultrastrong coupling regime is 
accessible and the terms beyond the RWA cannot be neglected. 

The interaction between a two-level system and a single mode of
the quantized electromagnetic field is described by the 
time-dependent Hamiltonian
($\hbar=1$)~\cite{micromaser}
\begin{equation}
  \begin{array}{c}
{\displaystyle
H(t)=H_0+H_I(t),
}
\\
\\
{\displaystyle
H_0=-\frac{1}{2}\,\omega_a \sigma_z +
\omega\left(a^\dagger a +\frac{1}{2}\right),
}
\\
\\
{\displaystyle
H_I(t)=f(t)\,[\,g \,\sigma_+\,(a^\dagger+a),
+g^\star \sigma_-\,(a^\dagger+a)\,],
}
\end{array}
\label{eq:noREWAquantum}
\end{equation}
where $\sigma_i$ ($i=x,y,z$) are the Pauli matrices,
$\sigma_\pm = \frac{1}{2}\,(\sigma_x\mp i \sigma_y)$
are the rising and lowering operators for the two-level system:
$\sigma_+ |g\rangle = |e\rangle$,
$\sigma_+ |e\rangle = 0$,
$\sigma_- |g\rangle = 0$,
$\sigma_- |e\rangle = |g\rangle$;
the operators $a^\dagger$ and $a$ create
and annihilate a photon:
$a^\dagger |n\rangle=\sqrt{n+1}|n+1\rangle$,
$a |n\rangle=\sqrt{n}|n-1\rangle$,
$|n\rangle$ being the Fock state with $n$ photons.
We assume sudden switch on/off of the coupling:
$f(t)=1$ for $0\le t \le \tau$, $f(t)=0$ otherwise.
For simplicity's sake, we
consider the resonant case ($\omega=\omega_a$) and
the coupling strength $g\in\mathbb{R}$.
The RWA is obtained when we neglect the term
$\sigma_+ a^\dagger$, which simultaneously
excites the two-level system and creates a photon,
and $\sigma_- a$, which de-excites the
two-level system and
annihilates a photon. In this limit, Hamiltonian
(\ref{eq:noREWAquantum}) reduces to the Jaynes-Cummings
Hamiltonian \cite{micromaser}, with a switchable coupling.
We set $\omega=1$, so that in the RWA the interaction time needed to transfer 
an excitation from the qubit to the field or vice versa 
($|e\rangle |0\rangle\leftrightarrow |g\rangle |1\rangle$)
is $\tau=\pi/2g$.

We consider as initial condition the state 
$|\Psi_0\rangle=|g\rangle|0\rangle$, i.e. the tensor product of the 
ground state of the two-level system and of the 
oscillator vacuum, so that within RWA such state is
the ground state of the overall system and therefore its dynamical
evolution must be ascribed to the terms beyond the RWA.  
We compute numerically the qubit-field state after the 
interaction time:
$|\Psi(\tau)\rangle = c_g(\tau)|g\rangle|\phi_g(\tau)\rangle+
c_e(\tau)|e\rangle|\phi_e(\tau)\rangle$, where
$|\phi_g\rangle$ and $|\phi_e\rangle$
are normalized states of the field.

\begin{figure*}[ht]
\includegraphics[angle=0.0, width=16cm]{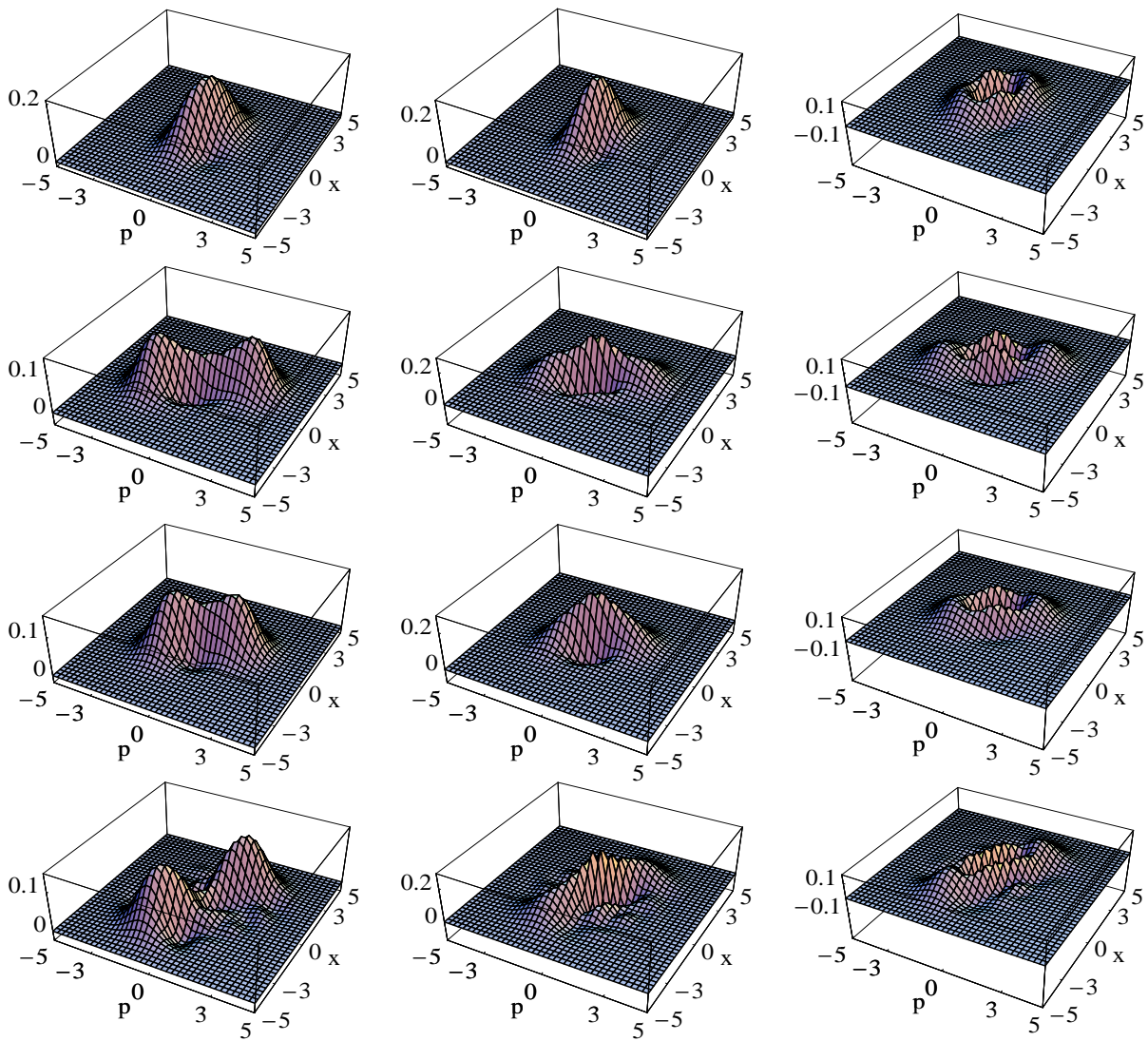}
\caption{Wigner function for different values of the coupling
strength $g$ and of the interaction time $\tau$:
$g=0.5$, $\tau=\pi/2g$ (first row), 
$g=1.5$, $\tau=\pi/2g$ (second row), 
$g=1$, $\tau=0.75 \pi/2g$ (third row), and
$g=1$, $\tau=1.5\pi/2g$ (fourth row), for the unconditional 
state $\rho$ (first column) and the conditional states
$\rho_g$ (second column) and $\rho_e$ (third column).} 
\label{fig:wigner}
\end{figure*}

By changing the interaction strength $g$ and the 
interaction time 
$\tau$ we can generate a great variety of states of the 
field, both in the unconditional case 
and in the conditional
case in which the final qubit state is measured, for 
instance in the $\{|g\rangle,|e\rangle\}$ basis. 
In the first case, the field state reads
$\rho={\rm Tr}_q |\Psi\rangle\langle\Psi|=
|c_g|^2|\phi_g\rangle\langle\phi_g|
+|c_e|^2|\phi_e\rangle\langle\phi_e|$, where
${\rm Tr}_q$ denotes the trace over the qubit subsystem;
in the latter case, we obtain the (pure) states
$\rho_g=|\phi_g\rangle\langle\phi_g|$ or 
$\rho_e=|\phi_e\rangle\langle\phi_e|$.
In Fig.~\ref{fig:wigner} we show the Wigner function
$W(x,p)$, with $x$ and $p$ position and momentum operators 
for the harmonic oscillator, 
at $g=0.5$, $g=1$, and $g=1.5$ and for different values of 
$\tau$, both for the unconditional state
$\rho$ and for the conditional states 
$\rho_g$ and $\rho_e$. 
Note that in all these instances, even at $g=0.5$
(first row of Fig.~\ref{fig:wigner}), the field 
is not simply in a squeezed state, since there are negative
components of the Wigner function. At largest interaction
strengths ($g=1.5$ in the second and $g=1$ in the
third and fourth rows of Fig.~\ref{fig:wigner})
we obtain exotic states of the field, very different from 
the squeezed 
states usually associated 
\cite{bastard,noriRMP} with the dynamical Casimir effect.   

\begin{figure}
\includegraphics[angle=0.0, width=8.5cm]{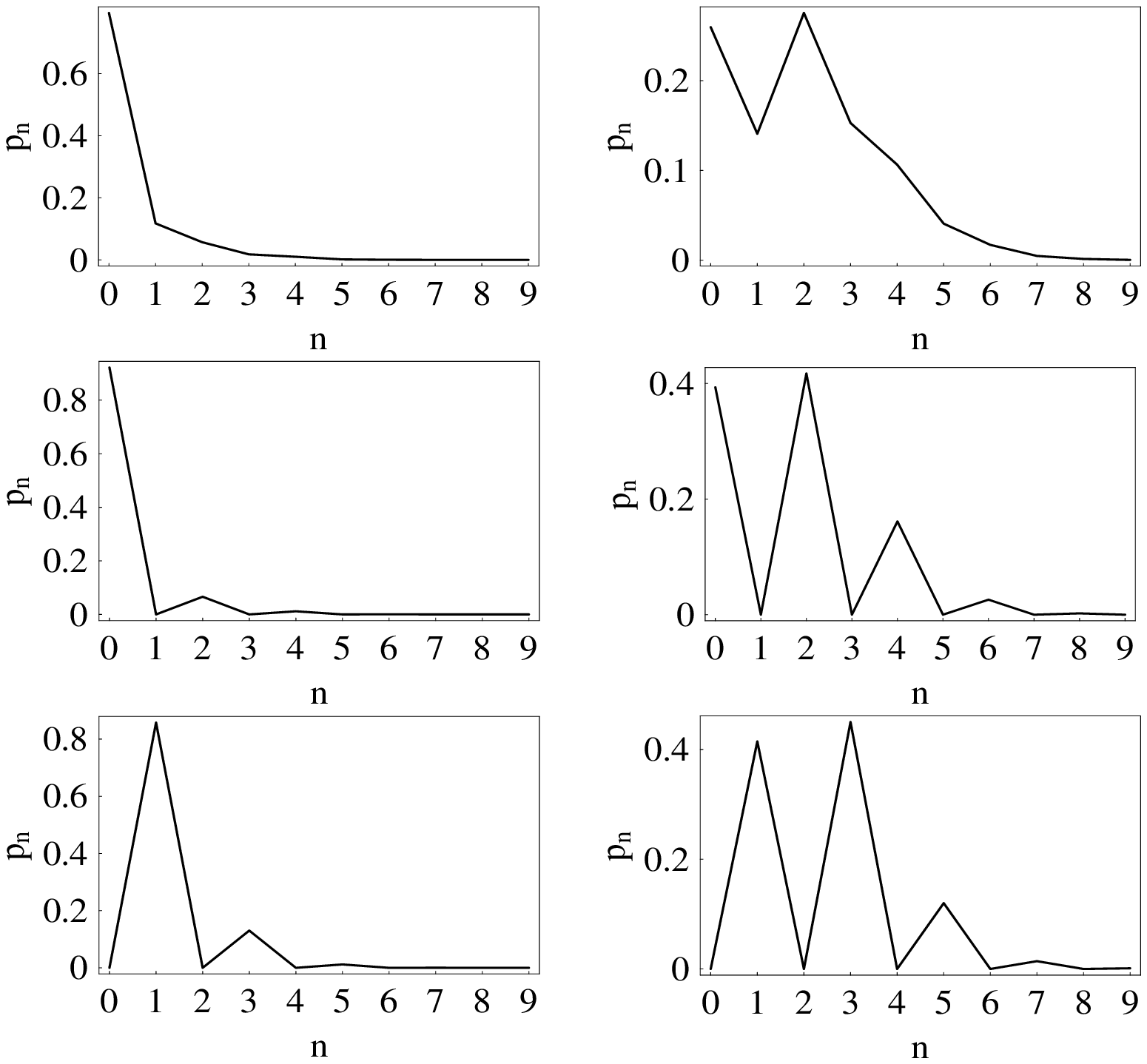}
\caption{Populations of the final states 
$\rho$ (first row), $\rho_g$ (second row), and 
$\rho_e$ (third row) of the field, 
for $\tau=\pi/2g$,
$g=0.5$ (left column) and $g=1.5$ (right column).}
\label{fig:populations}
\end{figure}

In Fig.~\ref{fig:populations} we show the populations 
$p_i$ ($\rho=\sum_{i,j} \rho_{ij} |i\rangle\langle j|$,
and $p_i=\rho_{ii}$)
of the states $\rho$, $\rho_g$ and $\rho_e$ for 
$g=0.5$ and $g=1.5$, with $\tau=\pi/2g$.
An interesting feature of these plots is that  
for the conditional states $\rho_g$ and $\rho_e$ 
only the states with respectively an even and an odd number of 
photons are populated. This follows from the 
fact that Hamiltonian (\ref{eq:noREWAquantum}) conserves the parity
$\Pi=\sigma_z (-1)^{a^\dagger a}$ of the excitations. 
Fig.~\ref{fig:populations} shows a significant population of
states with a rather large number of photons, demonstrating
the relevance of the DCE for the parameter values here considered.

\begin{figure}
\includegraphics[angle=0.0, width=8cm]{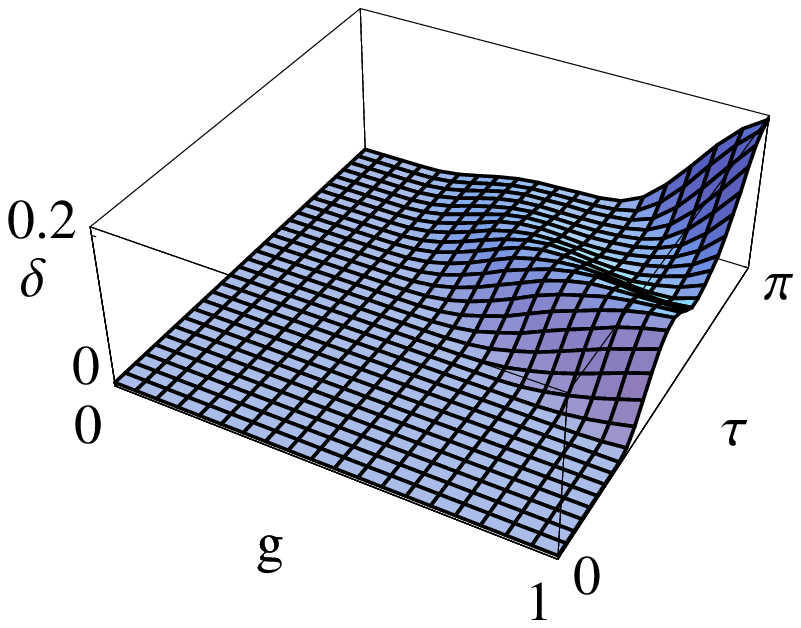}
\caption{Non-classicality indicator $\delta$ as a function of the
interaction strength $g$ and of the interaction time 
$\tau$ for the unconditional field state $\rho$.}
\label{fig:negativity}
\end{figure}

The non-classicality of the generated field states 
is characterized by the negativity parameter \cite{zyczkowski}
\begin{equation}
\delta=\int\int [\,|W(x,p)|-W(x,p)\,]\,dx\,dp.
\end{equation}
Note that such parameter vanishes for positive-definite 
Wigner functions, as it is the case for squeezed states. 
Therefore, a non-zero negativity distinguishes the states 
of Fig.~\ref{fig:wigner} from the usual squeezed states 
generated by parametric amplifiers \cite{noriRMP}. 
As shown in Fig.~\ref{fig:negativity}, negativity 
can be increased by either increasing the 
coupling strength $g$ or the interaction time $\tau$.
An interesting feature of this figure is that negativity 
becomes significantly different from zero only after a
finite interaction time $\tau$. The stronger the 
coupling strength $g$, the smaller is the interaction 
time requested to obtain non-classical states. 

\begin{figure}
\includegraphics[angle=0.0, width=8.5cm]{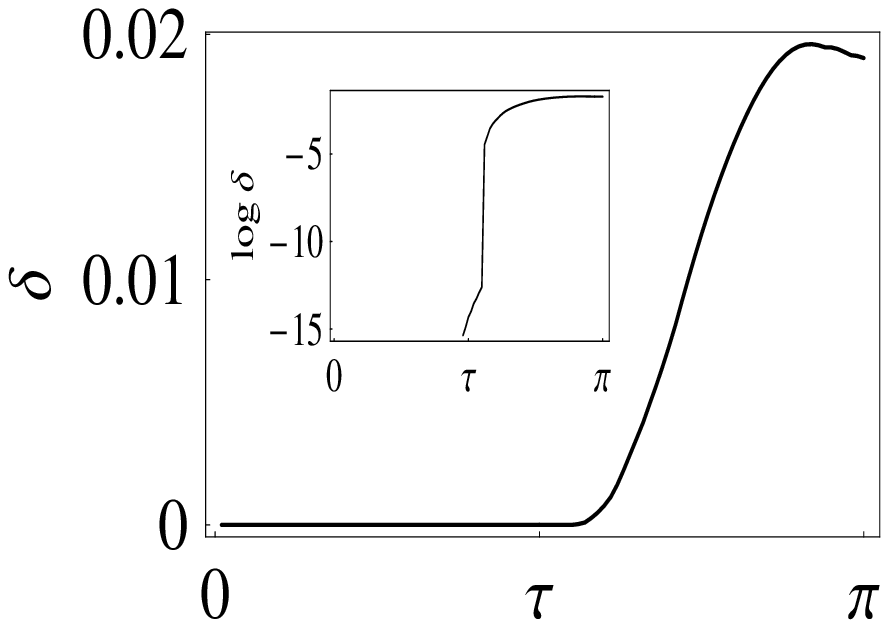}
\caption{Transition to non-classical states. 
Note the linear scale in the main plot and the logarithmic scale 
in the inset (over 12 orders of magnitude) for the
non-classicality indicator $\delta$.}
\label{fig:transitions}
\end{figure}

As shown in Fig.~\ref{fig:transitions}, there are signatures of
a sharp transition to non-classical states, $\delta>0$, 
i.e. with negative components of
the Wigner function, for instance at $\tau=\tau_c\approx 0.56 \pi$ for
$g=0.4$.
A similar behavior is observed also when the
parameter $\delta$ is drawn as a function of $g$ for 
a given $\tau$. 
Note that, by decreasing $\tau$ in 
Fig.~\ref{fig:transitions}, the transition at  $\tau=\tau_c$ 
leads to a drop in the parameter $\delta$ down to a value 
smaller than $10^{-15}$.
Hence, for practical purposed at $g=0.4$  the Wigner function
can be considered as non-negative for 
$\tau< \tau_c$, 
even though we cannot exclude that other transitions to states with 
negligible negativity occur at smaller values of $\tau$.   


The results of Figs.~\ref{fig:negativity}
and \ref{fig:transitions} can be interpreted by means
of the time-dependent perturbation theory, starting from the
initial state $|g\rangle|0\rangle$ with both the qubit and the
field in the ground state. The zeroth order of the Dyson series
is in agreement with the RWA result, that is no transitions occur  
to other states. 
The state $|e\rangle |1\rangle$ is coupled to $|g\rangle|0\rangle$
to first order. The Wigner function of the single-photon state 
$|1\rangle$ exhibits strong negativity. However, its weight 
in the Dyson series is 
not sufficient to overcome the positivity of the vacuum state $|0\rangle$.
To second order, the state $|g\rangle|2\rangle$ is excited.
Such states generates interference terms with $|g\rangle|0\rangle$,
which are responsible of the appearance of negative components 
in the Wigner function. For $g=0.4$, the second-order threshold
to obtain $\delta>0$ is $\tau=\tau_c^{(2)}\approx 0.60 \pi$, while a 
fourth-order calculation gives $\tau_c^{(2)}\approx 0.58 \pi$, in 
good agreement with the exact numerically computed result 
$\tau_c=0.56 \pi$. 

To summarize. we have shown that a sudden change of the coupling
constant in the ultrastrong coupling regime leads to the 
emergence of exotic states of the electromagnetic field, 
which cannot be described as squeezed or cat states. 
Such states are a manifestation of the dynamical Casimir
effect, whose first detectable consequence is the emission
of real photons. 
As analyzed in this work, these phenomena are present even 
if in the initial state both the qubit and the field are
in the ground state. 

It might appear surprising that, with such initial condition, 
one can measure final states with both the qubit and the 
field in an excited state \cite{vitiello} (see the third row of 
Fig.~\ref{fig:populations}). The energy for such excitations
comes from the setup that allows for a fast switching
of the interaction. For instance, when a two-level atom
enters a cavity, it experiences a braking force and is slowed down
\cite{cohen}.
The missing part of the kinetic energy is used to generate 
excited states of the atom and of the cavity. 

Since a nonadiabatic variation of the coupling constant 
in the ultrastrong coupling regime is a necessary condition
to perform fast quantum gates, the DCE appears as a generic
feature for high-speed quantum computation and communication protocols.
As a result, photons are emitted and 
the fidelity of quantum gates \cite{solano,noRWA}
or the capacity of quantum communication channels 
\cite{in_preparation_quantum_channel} deteriorate.
Therefore, the dynamical Casimir effect
can be the origin of a fundamental upper limit to the
maximum speed of quantum computation or communication protocols.

\begin{acknowledgments}
G.B. acknowledges the support by MIUR-PRIN project
``Collective quantum phenomena: From strongly correlated systems to
quantum simulators''.
\end{acknowledgments}


\end{document}